\documentclass[12pt,preprint]{aastex}
\usepackage{amsmath}
\shorttitle{rrlab}
\shortauthors{Ablimit \& Zhao.}

\begin{document}


\title{ The Milky Way's circular velocity curve and its constraint
on the Galactic mass with RR Lyrae stars}
\author{Iminhaji Ablimit\altaffilmark{1} and Gang Zhao\altaffilmark{1}}
%\email{$^\dagger$iminhajia@gmail.com}
\altaffiltext{1}{Key Laboratory of Optical Astronomy, National Astronomical Observatories,
Chinese Academy of Sciences, Beijing 100012, China; gzhao@nao.cas.cn, iminhaji@nao.cas.cn}
%\altaffiltext{2}{Department of Astronomy, Nanjing University, Nanjing 210046, China}
%\altaffiltext{3}{Key Laboratory of of Modern Astronomy and Astrophysics,
%Ministry of Education, Nanjing 210046, China}

%\date{}
%\pagerange{\pageref{firstpage}--\pageref{lastpage}} \pubyear{2007}
%\maketitle
%\label{firstpage}

\begin{abstract}
We present a sample of 1148 ab-type RR Lyrae (RRLab) variables
identified from Catalina Surveys Data Release 1, combined with SDSS
DR8 and LAMOST DR4 spectral data. We firstly use a large sample of
860 Galactic halo RRLab stars and derive the circular velocity distributions for
the stellar halo. With the precise distances and carefully determined radial velocities
(the center-of-mass radial velocities) by considering the pulsation of the RRLab stars in our
sample, we can obtain a
reliable and comparable stellar halo circular velocity curve. We
take two different prescriptions for the velocity anisotropy parameter $\beta$
in the Jeans equation to study the circular velocity curve and mass profile. We test two different solar peculiar motions in our calculation. Our
best result with the adopted solar peculiar motion 1 of (U, V, W) =
(11.1, 12, 7.2) km ${\rm s}^{-1}$ is that the enclosed mass of the Milky Way within 50 kpc is
$(3.75\pm1.33)\times10^{11}\,M_\odot$ based on $\beta=0$ and the
circular velocity $180\pm31.92$ (km $\, \rm s^{-1}$) at 50 kpc. This
result is consistent with dynamical model results, and it is also
comparable to the previous similar works.

\end{abstract}


\section{Introduction}

Deriving the rotation curve (RC) and mass distribution of every
substructures of the Milky Way enable us to understand the physical
process of galaxy formation. The RC of the Milky Way substructures
(i.e. bulge, thin/thick disc and halo) could directly constrain the
galactic mass, and could be used to study the local dark matter
content \citep{so01, bh13}. Many theoretical and observational works
have been investigating the RC and Galactic mass, but they are still
under debate and remain uncertain \citep{we10,ne13,m96,ib95}.

The most reliable method to derive the RC (or circular velocity curve) of
the Milky Way is by acquiring three-dimensional velocity
information. However, it is hard to measure reliable
three-dimensional velocities at larger distances and only radial
velocities are available with current telescopes. In the standard
approach, the star tracer population is assumed to be distributed
isotropically in the Galaxy, and the Jeans equation \citep{bi08} is
adopted for circular velocity at a given radius in the spherical
system. The radial velocity dispersion, number density and
velocity anisotropy parameter ($\beta$) of the Jeans equation are important
to the circular velocity. Xue et al.(2008) present the circular velocity curve and mass
of the Galaxy at $\sim 60$ kpc by using radial velocity dispersion
of 2041 blue horizontal branch (BHB) stars selected from SDSS DR6
and the Jeans equation with two constants of $\beta$ ($\beta=0 \,
\rm{and} \, 0.37$). Jeans equation and analytical models of the
phase-space distribution function of the tracer population have been
adopted for the circular velocity curve and mass estimate of the Milky Way at different
distances \citep{gn10,de12,ka12}. \cite{bh14} consider three
different prescriptions of $\beta$, one of them has a radially
varying $\beta$ given by ${\beta}(r) = (1+ {{r_a}^2}/{r^2})^{-1}$
(see Binney \& Tremaine 2008), where ${r_a}$ is the  anisotropy
radius. They show the variation of RC within $\sim 200$ kpc with
different models. More recently, \cite{hu15} use the current direct
and indirect measured values of $\beta$ to model the RC out to $\sim
100$ kpc with red clump stars selected from the LAMOST survey and K
giants selected from the SDSS/SEGUE survey.

 We take $\beta=0$ according to the observation, and also
adopt the new function from the developed dynamical model of \cite{wi15} for the velocity anisotropy profile in this work (see below sections for more details). We use ab-type RR
Lyrae (RRLab) stars (fundamental-mode pulsators) by combining the
SDSS DR8 and LAMOST DR4 spectral data to construct the circular velocity curve. RR Lyrae
variables (RRLs) are low-metallicity stars that have evolved through
the main sequence and He-burning variable stars on the horizontal
branch of the color-magnitude diagram. At near-infrared wavelengths
the absolute magnitude deviation could be as low as 0.02 mag for the single one, which means $\sim$ 1\%
uncertainty in distance for an individual star (see Beaton et al.
2016 for details). To the whole population, the scatter of absolute magnitudes of Milky Way RRab's field and globular cluster stars may reach $\sim 0.1$ mag \citep{da13,da14}. Comparing with other stars like BHBs and K giants
used in previous studies, RRLab stars are brighter and can be
detected at large distances, and they can be used as standard
candles (for more details of distance uncertainty see
below sections) because of the narrow luminosity-metallicity ($M_{\rm V}$ -
$[\rm Fe/H]$) relation in the visual band and
period-luminosity-metallicity relations in the near-infrared
wavelengths. RRL stars have been widely taken as a great tool for
determining the distances and studying the age, formation, and
structure of the Milky Way and local galaxies.

We select a sample of RRL stars which have measured radial velocity
and metallicity to investigate the circular velocity curve and mass
distribution of the Milky Way. We describe the sample selection of
RRLab stars in Section 2. We present the measured circular velocity
curve and the gravitational mass of the Milky Way in Section 3.
Section 4 contains our concluding remarks.

\section{The sample selection}
\label{sec:model}

Drake et al.(2013) analyzed 12227 RRLab stars ($\sim 9400$ are newly
discovered) from the 200 million public light curves in Catalina
Surveys Data Release 1. These stars span the largest volume of the
Milky Way ever surveyed with RRLs, covering $\sim 20,000\,
\rm{deg}^2$ of the sky (the equatorial coordinates shown in Figure 1). They
show data taken by the Catalina Schmidt Survey (CSS), the uncertainty in photometry is $\sim$0.03 mag
for V $<16$ mag and rises to 0.1 mag at V $\sim$19.2 mag (see Figure
3 of Drake et al. 2013). They also
identify RRLab stars with both of radial velocity and metallicity
information by combining Catalina photometry with Sloan Digital Sky
Survey (SDSS) spectroscopic data release 8 (DR8). Among the SDSS
matched stars, we selected 351 SDSS matched sample stars which have
both of velocity and metallicity in our work (in the region of
Galactic centric distance $\leq 50$ kpc).

The Large sky Area Multi-Object fiber Spectroscopic Telescope
(LAMOST) is a Chinese national scientific research facility operated
by National Astronomical Observatories, Chinese Academy of Sciences
\citep{z06,z12}. It is a special reflecting Schmidt telescope with
4000 fibers in a field of view of 20 $\rm{deg}^2$ in the sky. LAMOST
has completed its pilot survey which was launched in October 2011
and ended in June 2016. LAMOST DR4 has 7,681,185 low-resolution ($R
\sim 2000$) spectra in total, and we use these data to cross
match with the other CSS RRL stars (without SDSS matched samples),
within an angular distance of 3 arcsecond. We find 797 matched RRLab
stars,
%(145 stars observed several times \& 652 stars observed once by LAMOST),
and their metallicities and radial velocities are given in Figure 2.
For the LAMOST matched RRLab stars, we find that the metallicity
mean uncertainty is 0.19 dex (average metallicity is -1.1 dex), and
mean uncertainty of the radial velocity is 14.93 km$\,\rm s^{-1}$.
Comparing to the mean uncertainty
of the metallicity (0.1 dex, average metallicity is -1.38 dex) and radial velocity ($<15$
km$\,\rm s^{-1}$) of the SDSS matched RRLab
stars, the parameters derived LAMOST observations are reliable as well.
Totally, we have 1148 RRLab star samples which have precise distances and radial velocities
to study kinematics and mass of the Milky Way.

RRL stars are widely used for distance determination, and an
absolute magnitude-metallicity relation is the one which has been
used for distance calculation \citep{sa81}. One popular method is
adopted for the absolute magnitude given as (Chaboyer 1999; Cacciari
\& Clementini 2003),
\begin{equation}
M_{\rm V} = (0.23 \pm 0.04)([\rm Fe/H] + 1.5) + (0.59 \pm 0.03),
\end{equation}
where [Fe/H] is the metallicity of an RR Lyrae
star. Considering the uncertainties from the photometric calibration and the variations in metallicity
and uncertainty in RRab absolute magnitudes (also see Dambis et al. (2013) for the uncertainty in absolute magnitude), the overall uncertainties is around 0.15 mag. Corresponding to this, we derive a $\sim 7\%$ uncertainties in distances. For the heliocentric $d$ and
Galactocentric distances $R_{\rm GC}$, we use the equations,
\begin{equation}
d = 10^{(<V> - M_{\rm V} + 5)/5}\, {\rm kpc},
\end{equation}
where $<V>$ average magnitudes were corrected for interstellar medium extinction
using Schlegel et al. (1998) reddening maps, and
\begin{equation}
R_{\rm GC} = (R_\odot - d{\rm cos}\,b\, {\rm cos}\,l)^2 + d^2{\rm cos}^2\,b\, {\rm sin}^2\,l +
d^2{\rm sin}^2\,b \, {\rm kpc},
\end{equation}
where $R_\odot$, $l$ and $b$ are the distance from the sun to the Galactic center (8.33 kpc in this work, see Gillessen et al. 2009), Galactic longitude and latitude of the stars, respectively.

\section{Results and discussion}

Our sample of 1148 RRLab stars contains 288 thick disc stars with $1
< |z| < 4$ kpc, and also 860 halo stars with $|z| > 4$ kpc (see
Figure 3).

One of the important issues is deriving a dependable radial velocity
(the center-of-mass radial velocity) of the RRL stars. RRL stars are
well known to exhibit significant variation in radial velocity
measurements because of their pulsation (e.g., Liu 1991). Sesar
(2012) recently noted differences between velocities measured using
hydrogen and metallic lines as references and derived relationships
for correcting these. They found that the combination of three
Balmer lines would lead to uncertainties of a few km $\,\rm s^{-1}$.
Drake et al.(2013) combined the relationships given by Sesar (2012)
to produce an appropriate correction for the SDSS measurements.
Thus, we have the velocity in Galactic standard of rest frame with
corrected redial velocities of 350 SDSS matched halo RRLab stars within 50 kpc
directly from Drake et al. (2013). LAMOST observes every star three
times continuously with 30 minutes exposure each time, and some
objects are observed multiple times in the different observational
periods. (We have 145 multi-observed stars in our sample, and we
average all data of the multi-observed stars for stellar parameters
of them.) In addition, LAMOST derives the radial velocity by
averaging the velocities from the metallic lines, Balmer H$\alpha$,
H$\beta$ and H$\gamma$ lines. To accurately correct for velocity
variation, we need to know how the radial velocities were measured,
and the observed phase of the star. Using an average time of LAMOST
observations, the period and ephemeris of the RRLab, we obtain the
phase of observations for the each star, and keep the phase between
0.1 -- 0.95 because of the uncertain velocity correction out of that
region (see Sesar 2012). The correction method based on combinations
of Balmer and metallic lines with combined  relationships of Sesar
(2012) used in Drake et al. (2013) is adopted in this work as well.
The distribution of the original radial velocity and the
redetermined radial velocity by the proper correction is given in
Figure 4 (right panel). We find that the correction for the
pulsation velocities improves the velocity data quality with a mean
uncertainty of 14.43 km ${\rm s}^{-1}$. Our mean uncertainty has
good agreement with the uncertainties of Sesar 2012 (13 km ${\rm
s}^{-1}$) and Drake et al. 2013 (14.3 km ${\rm s}^{-1}$). We agreed
with the conclusion that the metallicity measurement is not affected
by the pulsation (Drake et al. 2013), and the distribution of
metallicities of halo tracers in our sample is shown in the left
panel of Figure 4. Therefore, we have reliable radial velocities of
510 stellar halo tracers with acceptable uncertainties from analysis of 1302
LAMOST spectra for further calculations.

To transform the heliocentric radial velocities ($V_{\rm h}$) of
the stars to the fundamental standard of rest (FSR) we
adopt the following equation by using the solar peculiar motion 1 of (U, V, W) =
(11.1, 12, 7.2) km ${\rm s}^{-1}$ (SM1, Binney \& Dehnen 2010) which are defined in a right-handed Galactic system with U
pointing toward the Galactic center, V in the direction of rotation,
and W toward the north Galactic pole. We adopt the solar peculiar motion 2 of (U, V, W) =
(11.1, 18, 7.2) km ${\rm s}^{-1}$ (SM2 has the only change in V, see Reid at al. 2014 and Rastorguev at al. 2017) for comparison.  We take a recent value of $235\pm7$ km ${\rm s}^{-1}$ for the local standard of rest ($\rm{V}_{\rm lsr}$, Reid at al. 2014 and Rastorguev at al. 2017) in the equation below,

\begin{equation}
V_{\rm FRS} = V_{\rm h} + {\rm U} {\rm cos}\, b\, {\rm cos}\, l + ({\rm V} + {\rm V}_{\rm lsr})
{\rm cos}\, b\, {\rm sin}\, l + {\rm W} {\rm sin}\, b.
\end{equation}
The calculation results are given in Figure 5. From the figure, it can be seen that the change of V in the solar peculiar motion slightly affect the distribution of $V_{\rm FRS}$.

%\subsection{RC of the thick disc}

%The 288 thick disc tracers distributed from $R_{\rm GC} = \sim 6$ kpc to $R_{\rm GC} = \sim 15$ kpc, we take 1 kpc as a bin. We have 37, 84, 81, 37, 15, 17, 9, 4 and 4 stars in the bins, respectively. We calculate the circular velocity and its uncertainty (considering the uncertainty of $V_{\rm GRS}$) of each star in the each bin, then make average of each bin, the equation is  (Binney \& Merrifield 1998),

%\begin{equation}
%V_{\rm C} = \frac{R}{R_\odot}(\frac{V_{\rm GRS}}{{\rm cos}\,b\, {\rm sin}\,l}),
%\end{equation}

%the distribution of $V_{\rm GRS}$ of thick disc tracers and circular velocity of the thick disc are shown in Figure 6. Because of the finite number, the derived thick disc
%rotation curve is not ideal and the uncertainty is large.

\subsection{\textbf{The circular velocity curve} by the halo tracers}

The sample has 860 halo tracers to the distance up to $R_{\rm GC}
\sim 50$ kpc. We take a simple and widely used method to derive the
circular velocity $V_{\rm C}$ by applying the velocity dispersion
$\sigma_r$ of tracers and the spherical Jeans equation (Binney \&
Tremaine 2008),

\begin{equation}
V^2_{\rm C} = -{\sigma}^2_r(\frac{d{\rm ln}\nu}{d{\rm ln}r} + \frac{d{\rm ln}{\sigma}^2_r}{d{\rm ln}r} + 2\beta),
\end{equation}
where $\sigma_r$, $\nu$ and $\beta$ are the Galactocentric radial
velocity dispersion, number density of tracer population and
velocity anisotropy parameter, respectively. A number of studies
follow a broken power-law distribution $\nu \propto r^{-\alpha}$,
and a shallow slope of $\alpha \sim 2-3$ up to a break radius
$r_{\rm b} \sim 16-27$ kpc and a steeper slope of $\alpha \sim
3.8-5$ out of $r_{\rm b}$ (Bell et al. 2008; Watkins et al. 2009;
Sesar et al. 2013; Faccioli et al. 2014). According to the number
density of RRL given by Watkins et al. 2009, we adopt $\alpha = 2.4$
for the inner halo ($r \leq 25$ kpc) and $\alpha = 4.5$ for the
outer halo ($r > 25$ kpc). We address $\sigma_r$ and $\beta$ below.

Before obtaining the Galactocentric radial velocity dispersion
$\sigma_r$, in the first step, we divide all stars into several bins
and calculate $\sigma_{\rm FRS}$ in each bin by averaging; the first
bin is to 10 kpc, and beyond 10 kpc we use 5 kpc to make a bin. We
take $\Delta\sigma_{\rm FRS} = (\sqrt{1/[2({\rm N} -
1)]})\sigma_{\rm FRS}$ for the uncertainty of $\sigma_{\rm FRS}$ in
each bin, where N is the number of members in tracer population in the bin. In the
first bin, for $<10$ kpc, we have 128 stars. There are 274, 176, 119, 76,
28, 24, 19 and 16 stars in the bins representing 10-15, 15-20, 20-25,
25-30, 30-35, 35-40, 40-45 and 45-50 kpc, respectively.  In the second
step, we derive the Galactocentric radial velocity dispersion
$\sigma_r$ at the Galactocentric distance $r$ for each bin by using
$\sigma_{\rm FRS}$ and the relation (Battaglia et al. 2005),

\begin{equation}
\sigma_{r} = \frac{\sigma_{\rm FRS}}{\sqrt{1 - \beta H(r)}},
\end{equation}
where

\begin{equation}
H(r) = \frac{r^2 + {R^2_\odot}}{4r^2} - \frac{(r^2 - {R^2_\odot})^2}{8r^3R_\odot}{\rm ln}{\frac{r + R_\odot}{r - R_\odot}},
\end{equation}
The velocity anisotropy parameter $\beta$ is described as,
\begin{equation}
\beta(r) = 1 - \frac{\sigma^2_t}{2{\sigma}^2_r} ,
\end{equation}
where $\sigma_t$ is the transverse velocity dispersion. Because of a lack of the proper motions for tracers, $\beta$ is usually taken to be a constant or taken from the modulated relations as discussed in Section 1. Deason et al.(2013) show the halo is isotropic $\beta = 0.0^{+0.2}_{-0.4}$ at $r = 26\pm6$ kpc.  Kafle et al.(2014) find $\beta = 0.4\pm0.2$ beyond 50 kpc.  For the $\beta$, we adopt two different scenarios; one is regarding it as a constant $\beta = 0$ out to 50 kpc, and the other is following  a new relation between $\beta$ and Galactocentric distance $r$ given by Williams \& Evans(2015), as,
\begin{equation}
\beta(r) = \beta_0 + \frac{(\beta_1 - \beta_0)r}{r + r_0} ,
\end{equation}
where $\beta_0 = 0.05$, $\beta_1 = 0.82$ and $r_0 = 18.2$ kpc. By the relation, we get the $\beta$ changing from $\sim$0.32 to $\sim$0.67 at the distance from 10 kpc to 50 kpc.  The calculated Galactocentric radial velocity dispersion with $\beta = 0$ and two SM are given in Figure 6. There are $\sim 2$ km$\,\rm s^{-1}$ discrepancy  between $\sigma_{r}$ at 50 kpc calculated with SM1 and SM2. From the Figure we can see that the Galactocentric radial velocity dispersion profile clearly declines in the range of $10 < r \leq 25$ kpc; the profile is slightly changed in the range $25 < r \leq 50$ kpc. Xue et al. (2008) fit an exponential function to their SDSS DR6 sample which is a linear approximation that gives $\sigma_{r} \approx 111 - 0.31r$. Gnedin et al.(2010) find a relation $\sigma_{r} \approx 120 - 0.22r$ in the range $25 < r < 80$ kpc. Their linear profile is less steep than Battaglia et al. (2005), $\sigma_{r} \approx 132 - 0.6r$. Compared to them, we get a different linear fit function in the range $25 < r \leq 50$ kpc with SM1 and $\beta=0$: $\sigma_{r} \approx 112 - 0.6r$; the uncertainty of the profile fit is dominated by the unclear parameters $\beta$ and $\alpha$. In order to express the profile better, we apply a power-law fit and find $\sigma_r \propto r^{\sim-0.38}$ with the solar peculiar motion 1 (the red dotted line in the Figure 6) \& $\sigma_r \propto r^{\sim-0.37}$ with the solar peculiar motion 2 (the black solid line in the Figure 6). The changes in the solar basic data (U, V, W and $V_{\rm lsr}$) can affect the velocity distributions and final results (also see Bhattacharjee et al.(2014)).

Finally, we can derive the circular velocity curve from the Jeans equation with values of $\nu$ and $\sigma_r$, and evaluate $\beta$ in two different ways for the halo tracers. The results with the solar peculiar motion 1 (hereafter results with SM1 are shown) are presented in Figure 7. This figure shows that the circular velocity and its uncertainty have higher values with a constant $\beta = 0$ than the varying $\beta$ of Williams \& Evans(2015).
$\beta$ changes from 0.32 to 0.67 based on the relation of Williams \& Evans(2015). This value is much higher than the isotropic one. We have $156.87\pm28.26$ km$\,\rm s^{-1}$ with varying $\beta$ and $180.0\pm31.92$ km$\,\rm s^{-1}$ with $\beta = 0$ at $R_{\rm GC} = 50$ kpc. Drake et al.(2013) discussed that the CSS RRLab stars are located across the Sagittarius stream. We have a clear increase in the circular velocity distribution around 40 kpc which is probably caused by the stream effect. We give two best simulated rotation curves based on results of Jeans equation with two different assumptions of $\beta$. Comparing with previous works, our two different circular velocity curves have a similar gentle declination and comparable circular velocity curve with moderate uncertainty. The isotropic one seems more reasonable than the varying $\beta$ within 50 kpc. It can be determined that the circular velocity and its uncertainty depend on the adopted values of parameters like $\nu$ and $\beta$ in the Jeans equation. We also need to measure all three dimensional velocity information and study the Galaxy with a non-spherical method. Newly developed telescopes can enable us to achieve that purpose in the future.

\subsection{Mass estimate}

The total mass and mass profile of the Milky Way affect the Milky Way's composition, structure, dynamical
properties and formation history. The circular velocity of the objects can be described by the relation between equality of the centripetal and gravitational force, as

\begin{equation}
\frac{m V^2_{\rm C}}{r} = m\frac{{\rm d}\Phi}{{\rm d}r}
\end{equation}
where $m$ and $\Phi$ are the mass of the object and the gravitational potential which satisfies the Poisson equation, respectively. Therefore, the following equation is simply derived,

\begin{equation}
\frac{m V^2_{\rm C}}{r} = \frac{{\rm G}mM(r)}{r^2}
\end{equation}
where $M(r)$ is the enclosed total mass within $r$.

We can thus present the mass profile estimate for the Milky Way by using the equation,
\begin{equation}
M(r) = {V^2_{\rm C} r}/{\rm G}.
\end{equation}
The crucial issue in this equation is how to derive the circular velocity $V_{\rm C}$ because proper motion measurements are incomplete and the mass estimate model with line-of-sight velocity measurements depends on several uncertain parameters. We adopt one of the two different ways to estimate the mass of the Milky Way, which only uses the line-of-sight velocity measurements. There have been many works computing the mass profile by using the fitting model with different tracer populations, producing the estimated mass profile within 50 kpc. Kochanek (1996) estimated a mass of the Galaxy within 50 kpc of $(4.9\pm1.1)\times10^{11}\,M_\odot$ by using the satellites of the Galaxy and the Jaffe potential model. Wilkinson \& Evans (1999) used 27 globular clusters and satellite galaxies with a Bayesian likelihood method and a spherical halo mass model to give the mass of $M(50 {\rm kpc}) = (5.4^{+0.2}_{-3.6})\times10^{11}\,M_\odot$, and a similar method used by Sakamoto et al.(2003) with the sample of 11 satellite galaxies, 137 globular clusters and 413 field horizontal-branch stars,  and they found the mass of $M(50 {\rm kpc}) = 1.8-2.5\times10^{11}\,M_\odot$. Later, Smith et al.(2007) considered three models based on a
sample of high-velocity stars from the RAVE survey, and the results from their three models are $(4.04^{+1.1}_{-0.76})\times10^{11}\,M_\odot$, $(3.87^{+0.64}_{-0.56})\times10^{11}\,M_\odot$ and $(3.58^{+0.04}_{-0.17})\times10^{11}\,M_\odot$ within 50 kpc. Interestingly, the results of Smith et al.(2007) perfect match with our result based on $\beta = 0$. Recently, Deason et al.(2012) find $M(50 {\rm kpc}) = \sim 4.0\times10^{11}\,M_\odot$ from about 4000 blue horizontal branch stars with the same method as Xue et al.(2008) and this paper.

We obtain $(3.75\pm1.33)\times10^{11}\,M_\odot$ and $(2.85\pm1.03)\times10^{11}\,M_\odot$ by deriving $180.0\pm31.92$ km$\,\rm s^{-1}$ based on the Jeans equation with $\beta = 0$ and $156.87\pm28.26$ km$\,\rm s^{-1}$ with varying $\beta$ within 50 kpc, respectively. Williams \& Evans(2015) develop a new dynamical model to study the circular velocity curve and the mass profile as a function of distance. They show the maximum likelihood mass profiles with Galactocentric distance and find the mass of $4.5\times10^{11}\,M_\odot$ at 50 kpc. Our result based on the relation of $\beta$ with distance found by Williams \& Evans(2015) is lower than the results of Williams \& Evans(2015) and some of the other works mentioned above. However, our result including the error with $\beta = 0$ and the solar peculiar motion 1 has a good fit with the results of the model and previous works. We plot this result in Figure 8 by marking it with the red word `A17'. As we can see in the figure, our result covers the results of the model and previous works.

\section{Concluding remarks}

We study the kinematics of 1148 RRLab stars to derive the circular velocity curve and mass profile of the Milky Way. We obtain the radial velocity dispersion out to 50 kpc by using measured radial velocity and metallicity profiles of 860 CSS RRLab halo stars from SDSS DR8 and LAMOST DR4. The spectral parameters of RRLab in this work show that results from the two survey projects are comparable. We consider the influence of the pulsation of RRLab stars on the radial velocity and the velocity uncertainty is reduced in a reasonable way. By measuring the precise Galactocentric distance ($\sim 7\%$ uncertainty) and the velocity dispersion, we adopt the parameter of the RRL star number density from a comprehensive study, and take two different anisotropy profiles to model the circular velocity curve and constrain the mass of the Milky Way. Two sets of V in the solar peculiar motion are considered in the calculation, and it is worth to note that the whole basic solar data have influence on the results. We derive our best result of $M(50 {\rm kpc}) = (3.75\pm1.33)\times10^{11}\,M_\odot$ by obtaining $180.0\pm31.92$ (km $\rm s^{-1}$) based on the Jeans equation with $\beta = 0$ and the solar peculiar motion 1 (SM1). We find a lower mass of $M(50 {\rm kpc}) = (2.85\pm1.03)\times10^{11}\,M_\odot$ with the circular velocity of $156.87\pm28.26$ (km $\rm s^{-1}$) by adopting SM1 and a new function of $\beta$ with distance given by a new developed dynamical model (Williams \& Evans 2015). More recently,  Bobylev et al.(2017) studied the halo with different Galactic gravitational potential models such as a spherical logarithmic Binney potential, a Plummer sphere and a Hernquist potential. The resulting Galactic masses within 50 kpc based on the three models are $(4.09\pm0.2)\times10^{11}\,M_\odot$, $(4.17\pm0.34)\times10^{11}\,M_\odot$ and $(4.17\pm0.32)\times10^{11}\,M_\odot$, respectively (see Bobylev et al. 2017). In conclusion, it is always very useful to see and compare the results of different objects (or/and different approaches) to the problem, and our result is in good agreement with results from theoretical and observational studies.

\acknowledgments This work was funded by China Postdoctoral Science Foundation(2015LH0015),
and also supported by the Natural Science
Foundation of China under grant number 11390371 and 11233004. We thank the anonymous referee for the useful comments. We thank Xue X. X., Zhao J. K., Liu C. and Yang F. for the useful discussions.
Guoshoujing Telescope (the Large Sky Area Multi-Object Fiber Spectroscopic Telescope LAMOST) is a National Major Scientific Project built by the Chinese Academy of Sciences. Funding for the project has been provided by the National Development and Reform Commission. LAMOST is operated and managed by the National Astronomical Observatories, Chinese Academy of Sciences.

%\bibliographystyle{apj}
%\bibliography{haji}

\begin{figure}
\centering
\includegraphics[totalheight=4.5in]{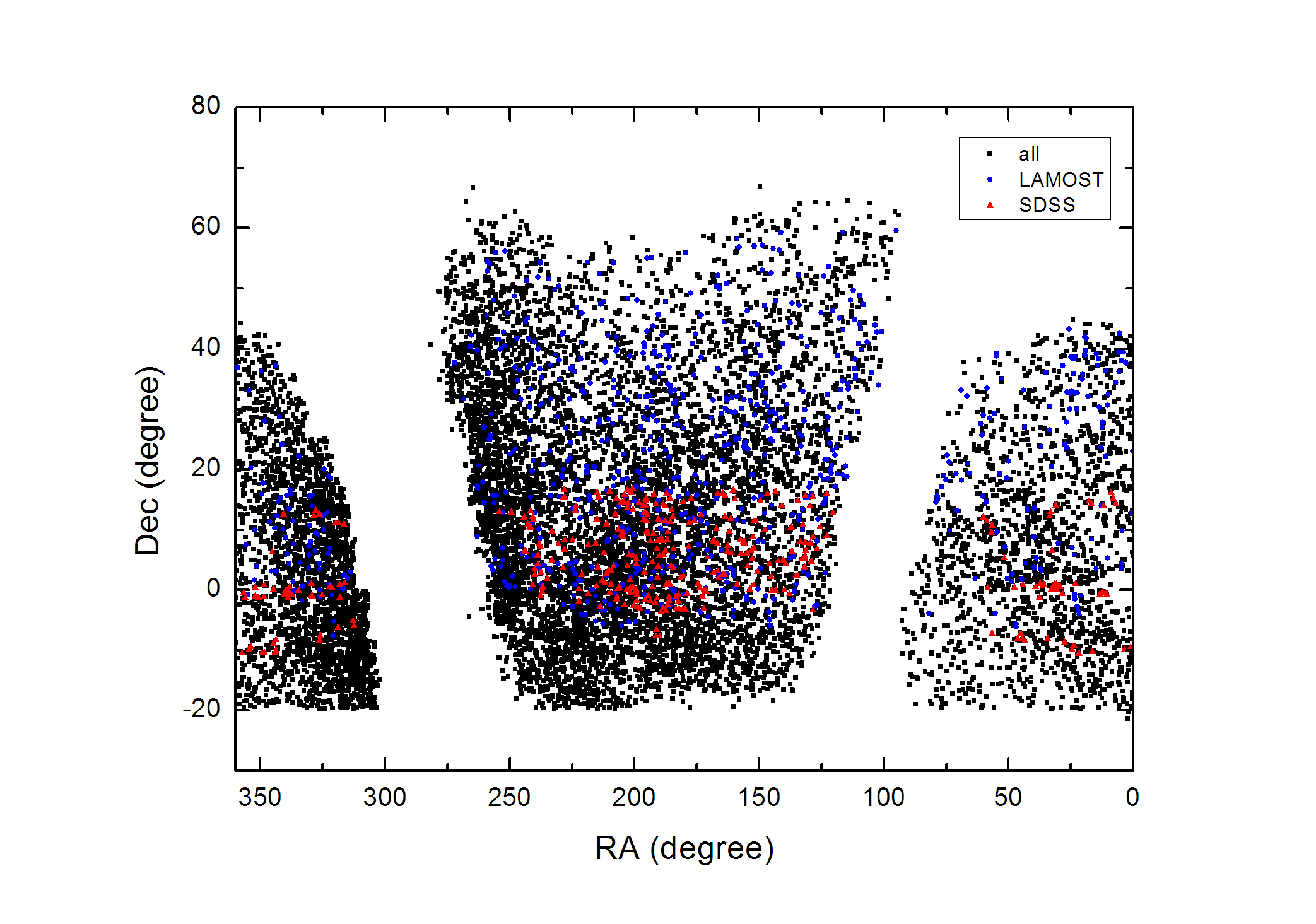}
\caption{The equatorial coordinates for all RRLab stars and our selected sample.}\label{fig:1}
\end{figure}

\clearpage

\begin{figure}
\centering
\includegraphics[totalheight=2.8in,width=3.2in]{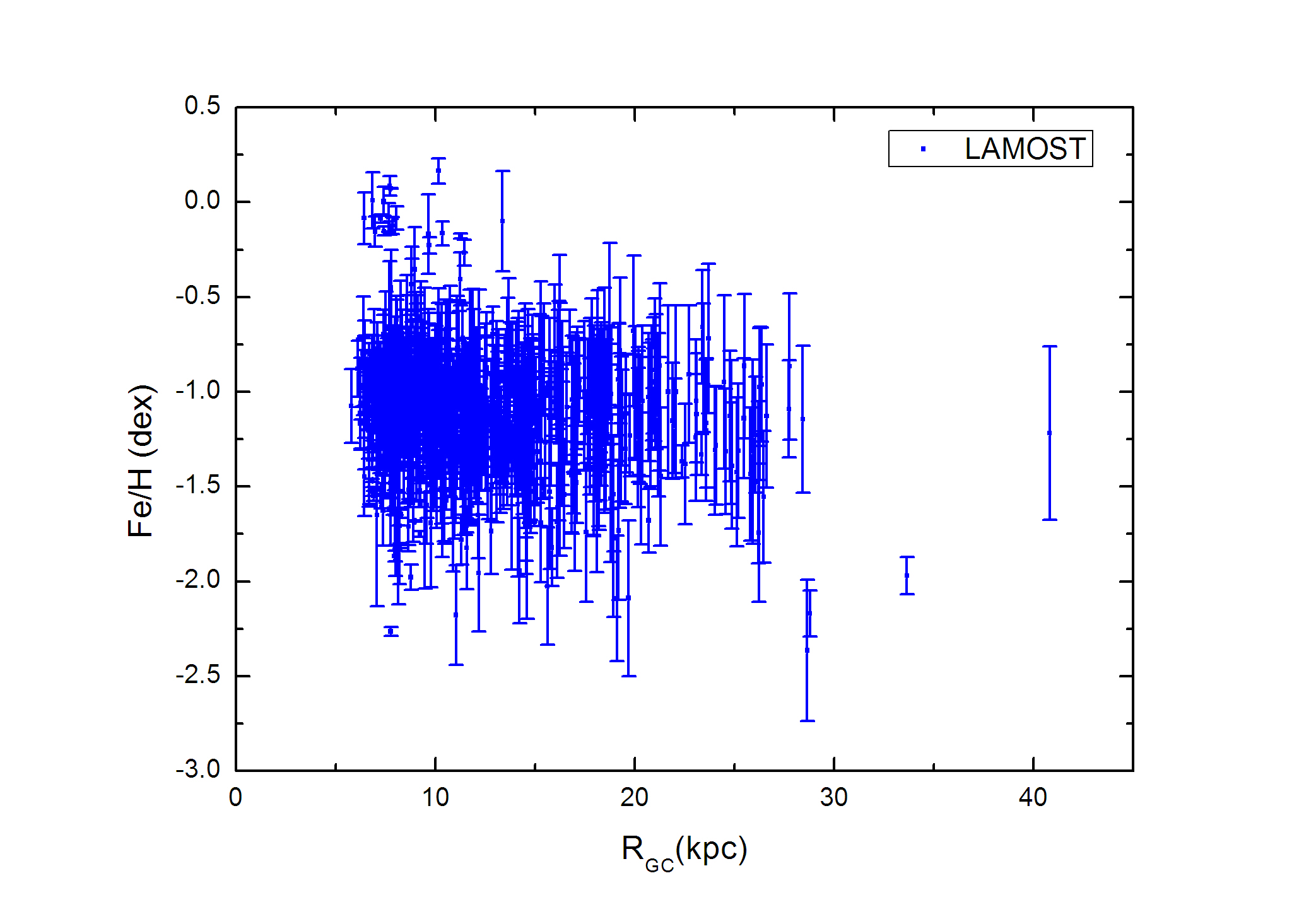}
\includegraphics[totalheight=2.8in,width=3.2in]{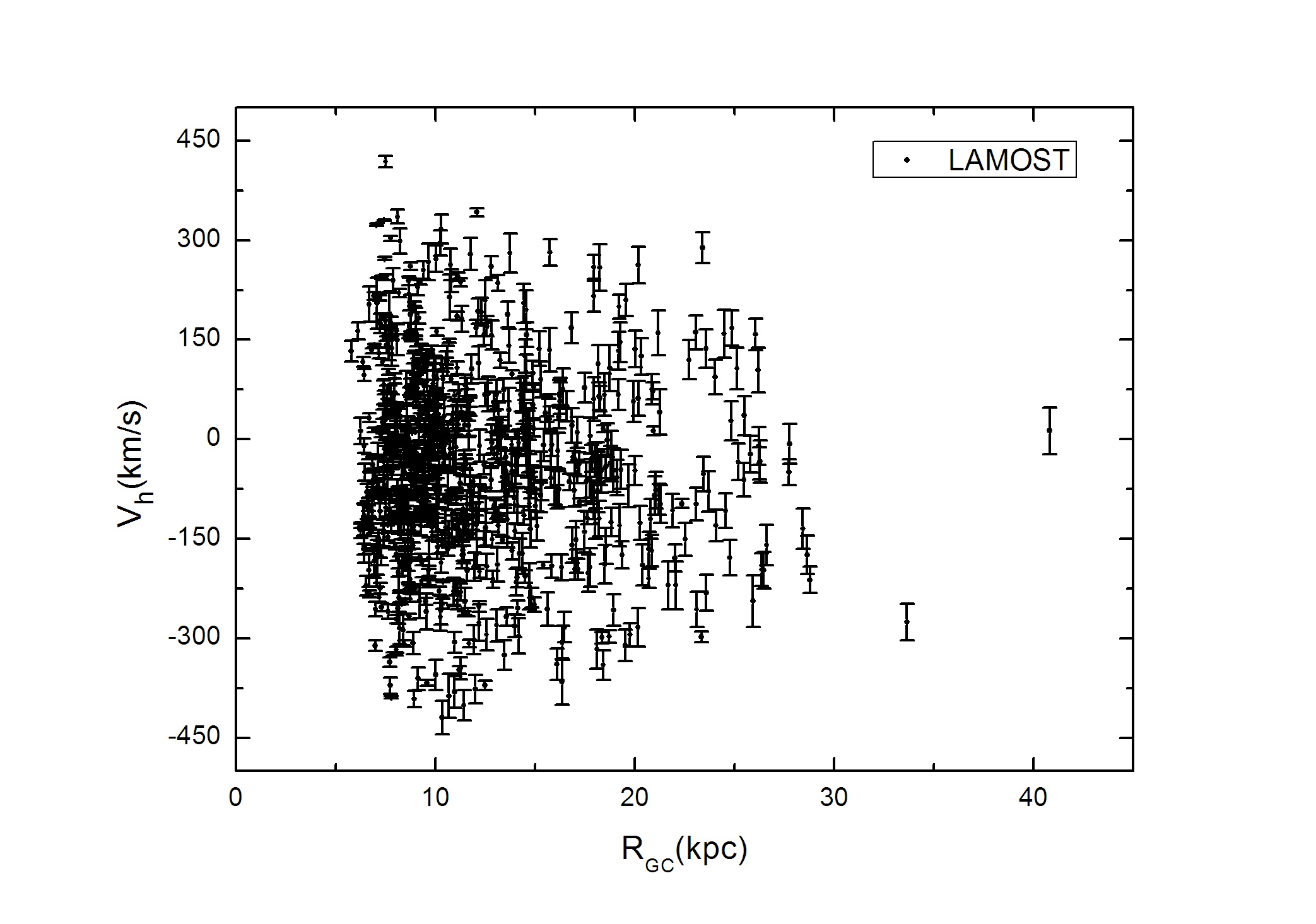}
\caption{The metallicity (left) and radial velocity (right panel) profiles of RRLab stars from LAMOST DR4.}\label{fig:1}
\end{figure}

\clearpage

%\begin{figure}
%\centering
%\includegraphics[totalheight=4.5in]{pic/f3}
%\caption{The metallicity distribution of our selected sample.}\label{fig:1}
%\end{figure}

%\clearpage

\begin{figure}
\centering
\includegraphics[totalheight=4.5in]{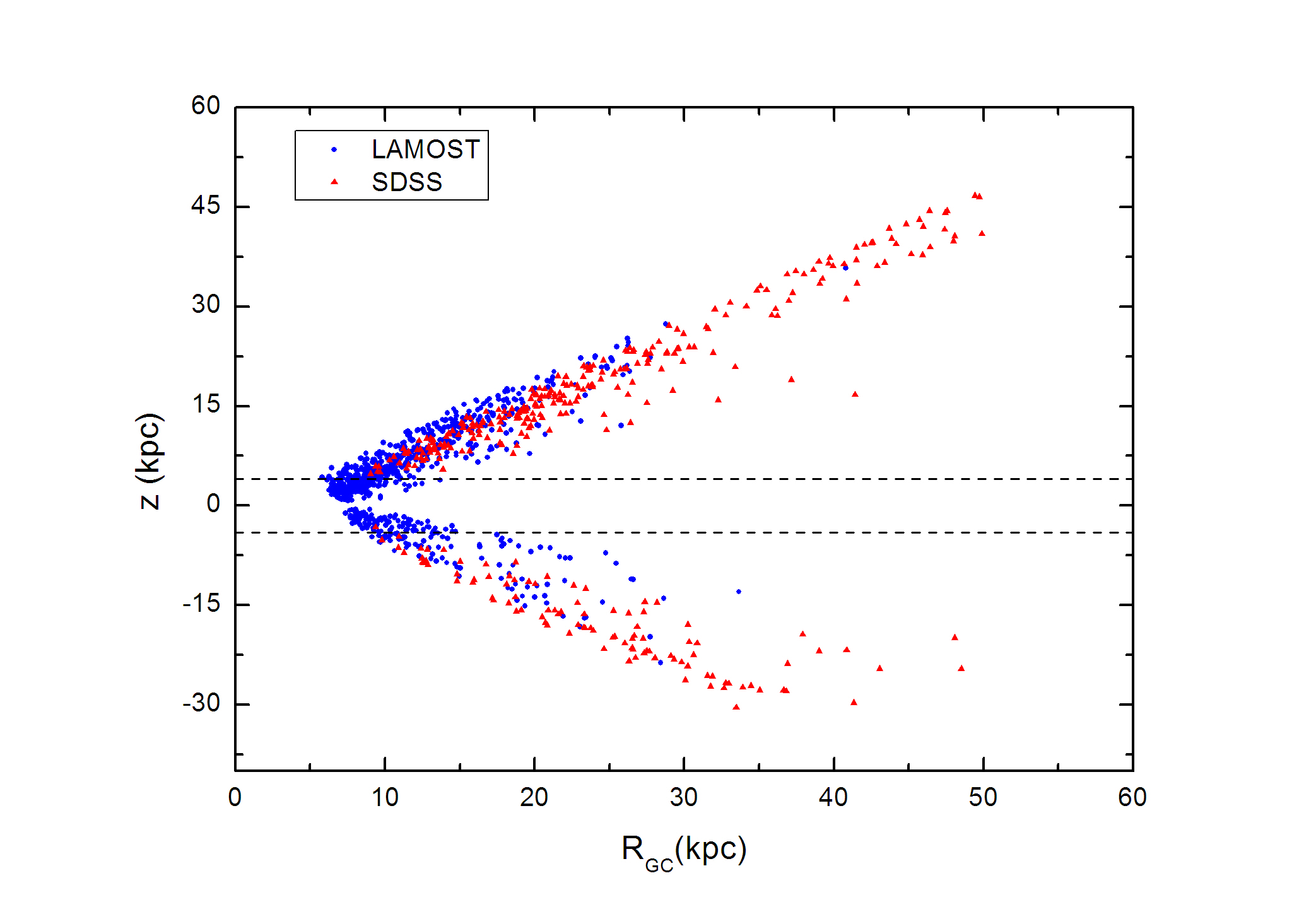}
\caption{Spatial distribution of whole sample in z-R plane, the
dotted lines are z= 4 kpc and z=-4 kpc, respectively.}\label{fig:1}
\end{figure}

\clearpage

\begin{figure}
\centering
\includegraphics[totalheight=2.8in,width=3.2in]{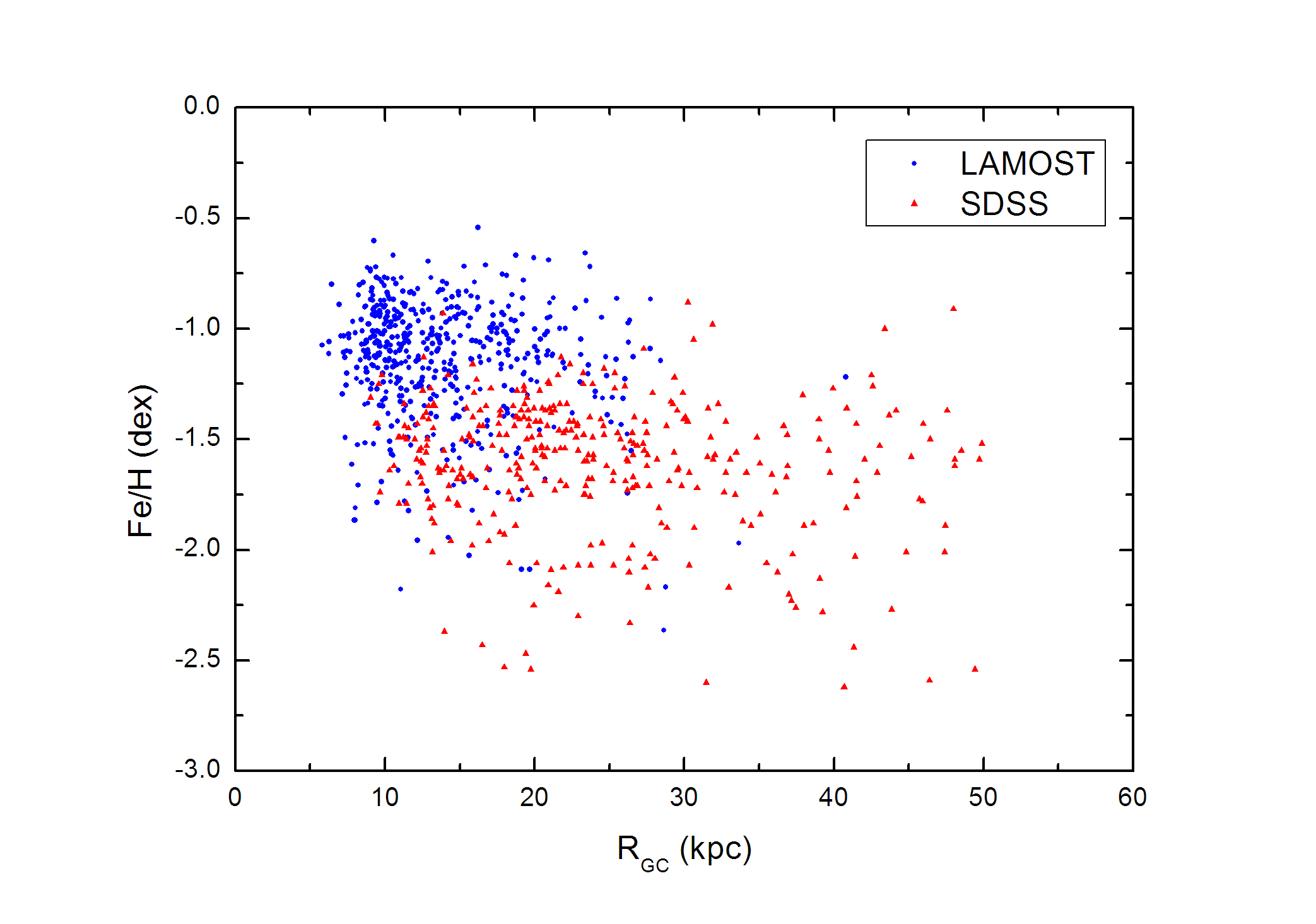}
\includegraphics[totalheight=2.8in,width=3.2in]{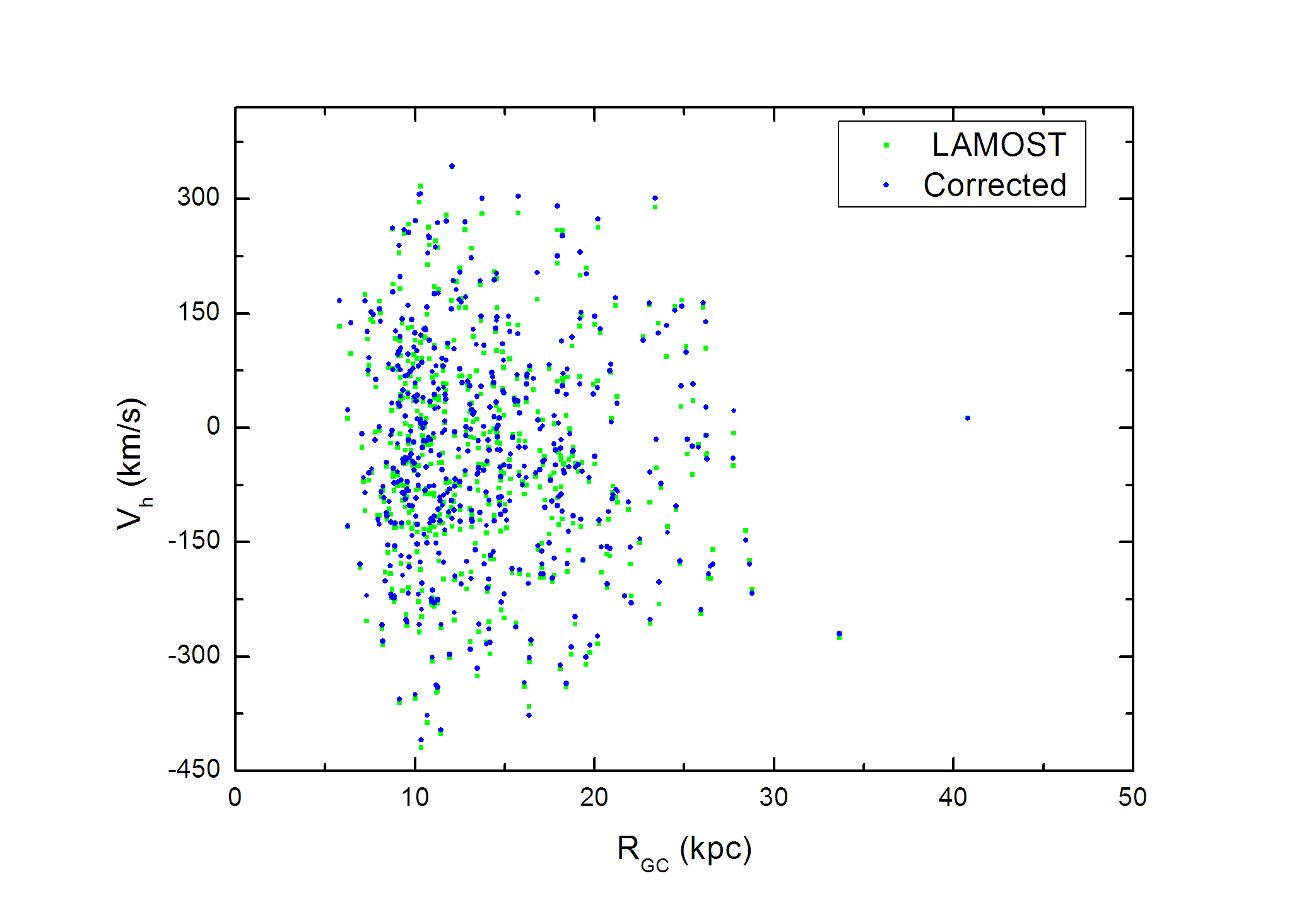}
\caption{The metallicity distribution (left panel) of all halo tracers. The right panel shows the heliocentric radial velocity distribution of LAMOST matched halo tracers, the green solid squares are the original values from LAMOST catalog, and the blue solid circles are the corrected values for the pulsation.}\label{fig:1}
\end{figure}

\clearpage

\begin{figure}
\centering
\includegraphics[totalheight=4.5in]{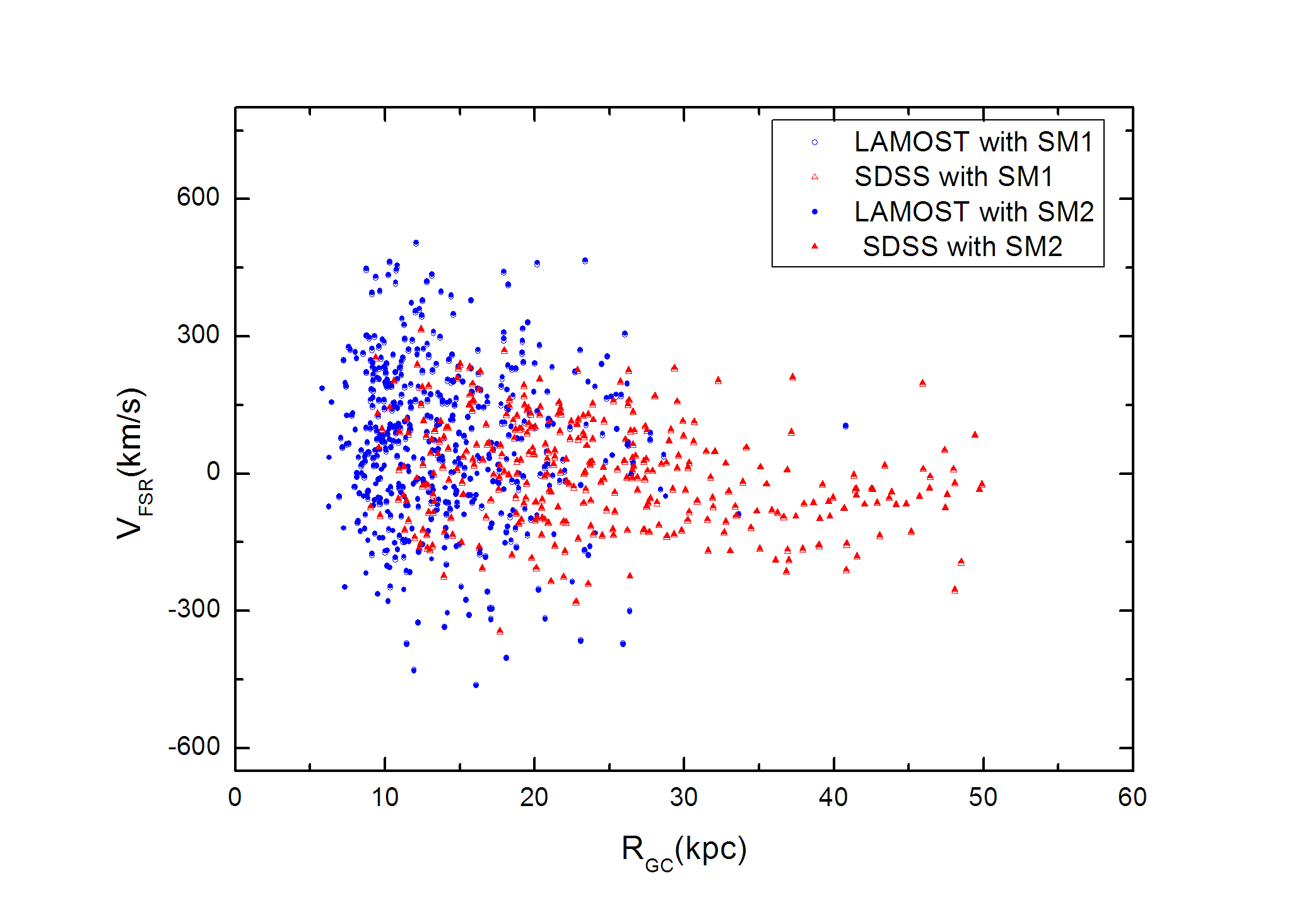}
\caption{$V_{\rm FSR}$ distributions of halo tracers. The open circle and triangle are based on the adopted solar peculiar motion 1, and the filled circle and triangle are based on the adopted solar peculiar motion 2}\label{fig:1}
\end{figure}

\clearpage

%\begin{figure}
%\centering
%\includegraphics[totalheight=2.8in,width=3.2in]{pic/f6a}
%\includegraphics[totalheight=2.8in,width=3.2in]{pic/f6b}
%\caption{ Distribution of $V_{\rm GSR}$ of the disc tracers(left) and circular velocity of the thick disc
%(right).}\label{fig:1}
%\end{figure}

\clearpage

\begin{figure}
\centering
\includegraphics[totalheight=4.5in]{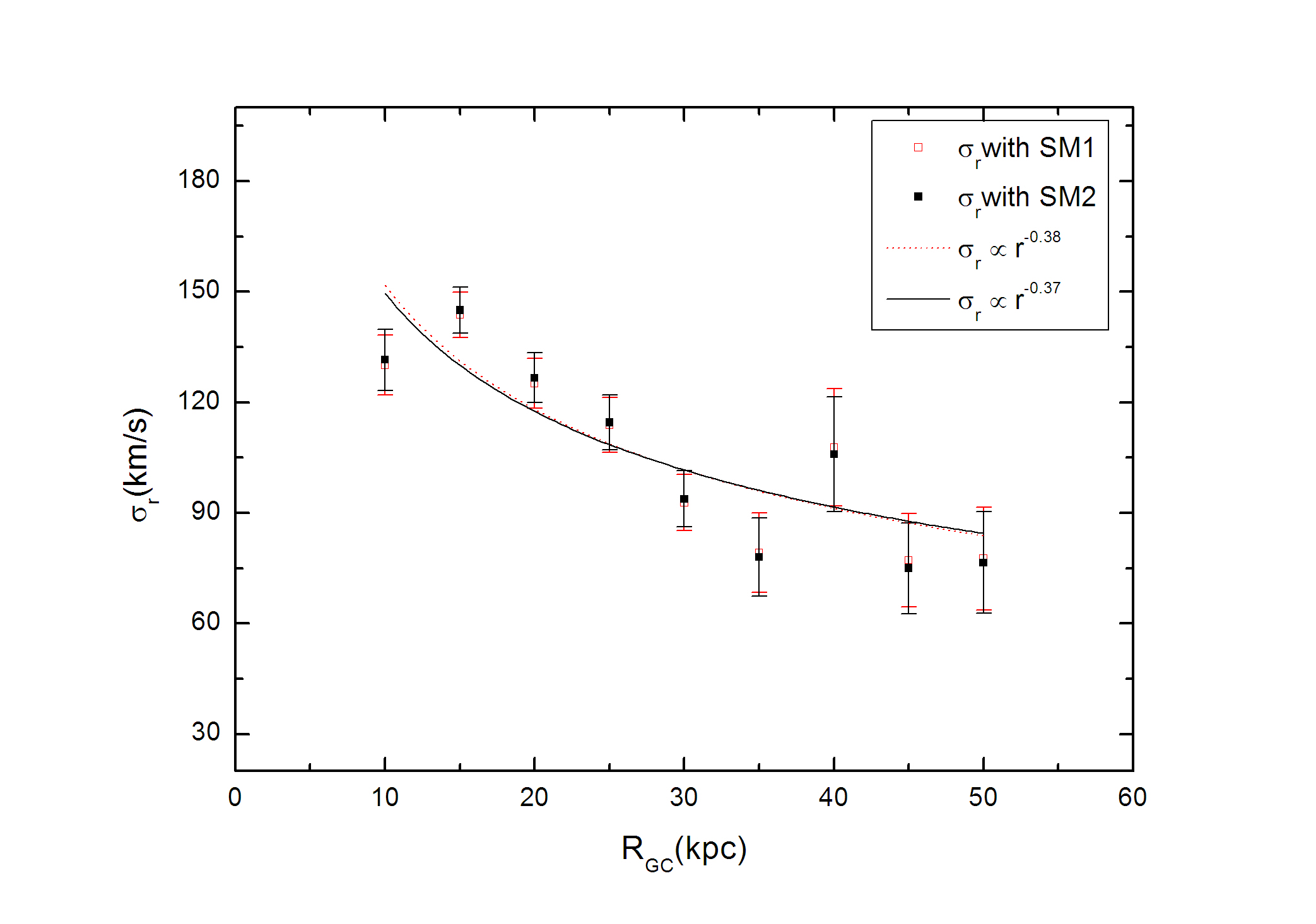}
\caption{The Galactocentric radial velocity dispersion $\sigma_r$ from halo tracers.}\label{fig:1}
\end{figure}

\clearpage

\begin{figure}
\centering
\includegraphics[totalheight=4.5in]{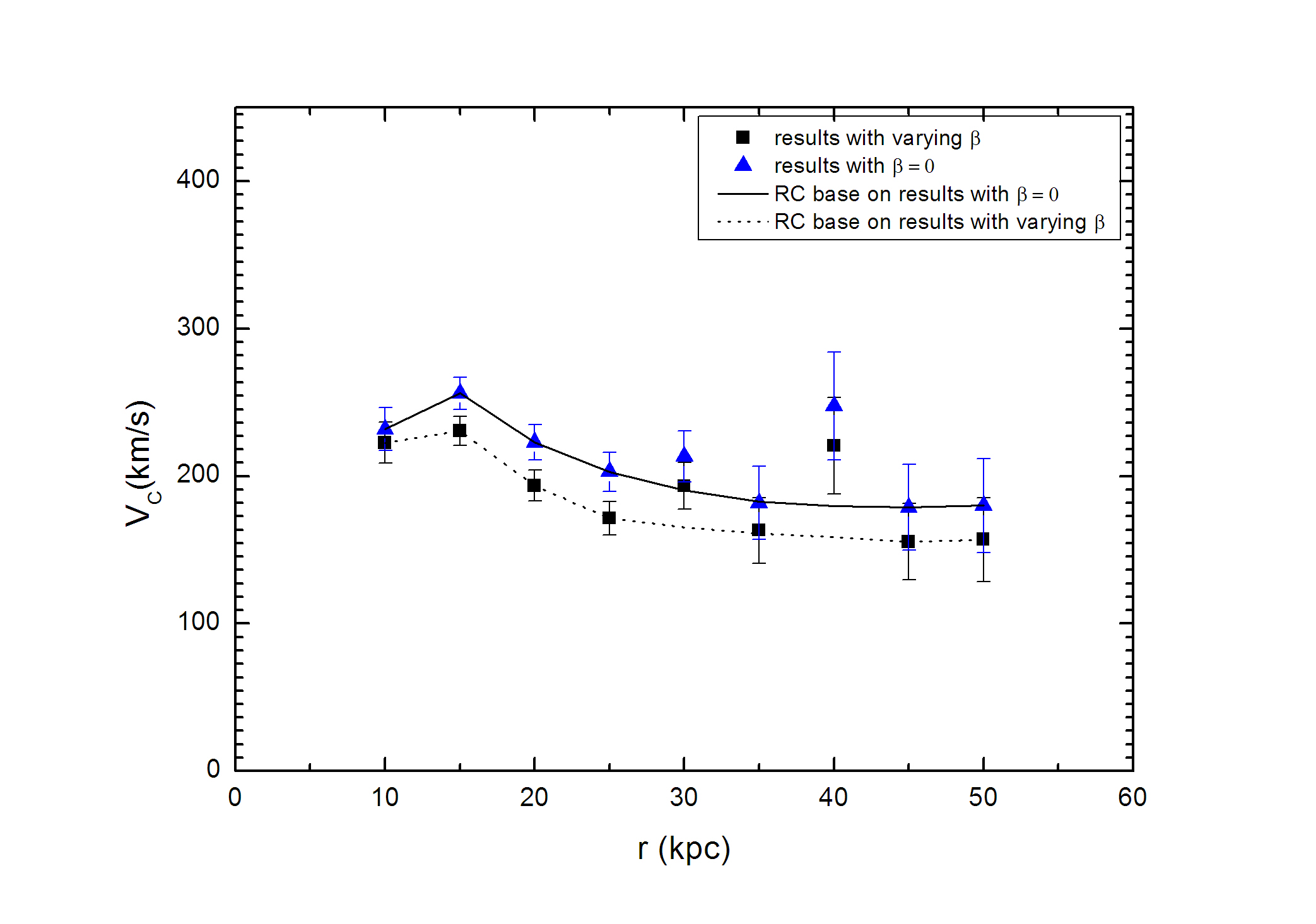}
\caption{Circular velocity distribution derived from halo
tracers (RC means circular velocity curve in the Figure). The blue triangle and solid line is resulted with $\beta = 0$, the black quadrangle and dotted line is calculated by varying $\beta$ with $r$ based on the relation given by Williams \& Evans(2015)}\label{fig:1}
\end{figure}

\clearpage

\begin{figure}
\centering
\includegraphics[totalheight=4.5in]{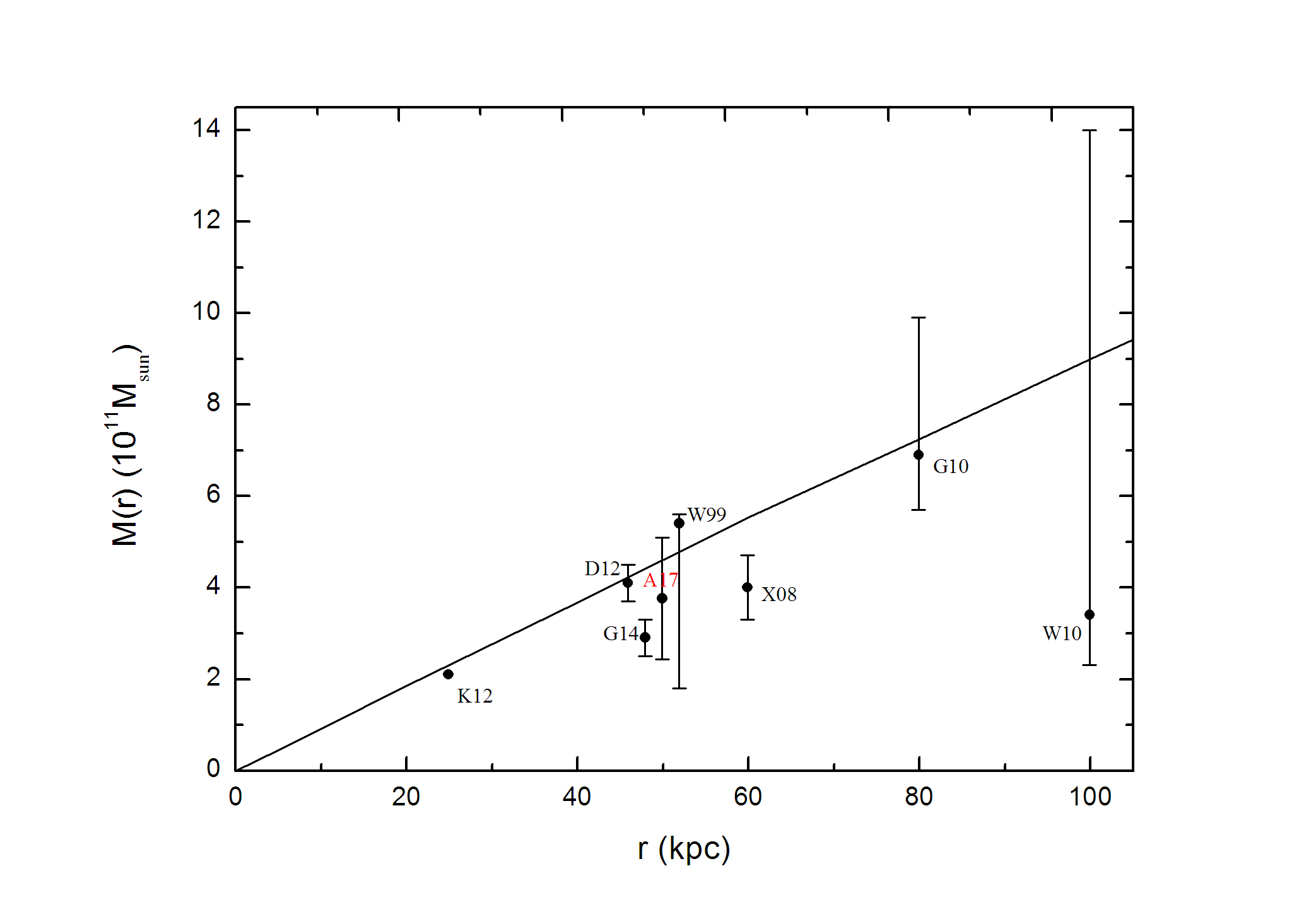}
\caption{The mass profile of the Milky Way. Our result with $\beta =0$ is marked by the red `A17'. The black line is from the maximum likelihood mass profiles of Williams \& Evans(2015). Other studies of the Milky Way cumulative mass distribution with error bars are plotted: K12 (Kafle et al. 2012), D12 (Deason et al. 2012), G14 (Gibbons et al. 2014), W99 (Wilkinson \& Evans 1999), X08 (Xue et al. 2008), G10 (Gnedin et al. 2010) and W10 (Watkins, Evans \& An 2010). To distinguish clearly our result from results of D12, G10, and W99 at 50 kpc, we slightly move the positions (r) of D12, G10, and W99.}\label{fig:1}
\end{figure}

\clearpage

\end{document}